\documentclass{aa}
\usepackage{graphicx}
\usepackage{natbib}
\bibpunct{(}{)}{;}{a}{}{,}
\begin{document}
\title{The effects of inclination, gravity darkening and differential
  rotation on absorption profiles of fast rotators}


\titlerunning{Inclination and gravity darkening in fast rotators}

\author{A. Reiners\inst{1}}

\offprints{A. Reiners}

\institute{Hamburger Sternwarte, Universit\"at Hamburg,
  Gojenbergsweg 112, 21029 Hamburg, Germany\\
  \email{areiners@hs.uni-hamburg.de}}

\date{accepted by A\&A}

\abstract{Mechanisms influencing absorption line profiles of fast
  rotating stars can be sorted into two groups; (i) intrinsic
  variations sensitive to temperature and pressure, and (ii) global
  effects common to all spectral lines. I present a detailed study on
  the latter effects focusing on gravity darkening and inclination for
  various rotational velocities and spectral types. It is shown that
  the line shapes of rapidly and rigidly rotating stars mainly depend
  on the equatorial velocity $v_{\rm e}$, not on the projected
  rotational velocity $v\,\sin{i}$ which determines the lines width.
  The influence of gravity darkening and spectral type on the line
  profiles is shown. The results demonstrate the possibility of
  determining the inclination angle $i$ of single fast rotators, and
  they show that constraints on gravity darkening can be drawn for
  stellar samples. While significant line profile deformation occurs
  in stars rotating as fast as $v_{\rm e} \ga 200$\,km\,s$^{-1}$, for
  slower rotators profile distortions are marginal. In these cases
  spectral signatures induced by, e.g., differential rotation are not
  affected by gravity darkening and the methods applicable to slow
  rotators can be applied to these faster rotators, too}

\maketitle

\section{Introduction}
\label{Introduction}

The information contained in line profiles of rotating stars has been
studied since the recognition of spectral Doppler broadening itself.
The multitude of different mechanisms influencing line profiles
presents a confusing picture for the possibility of determining them.
Approximations exist for rotational line broadening, (linear) limb
darkening, differential rotation, turbulence, rotational flattening
and gravity darkening, etc.. While distinguishing these interacting
mechanisms is a delicate issue, in the case of slow rotators
($v\,\sin{i} < 50$\,km\,s$^{-1}$) rotational flattening and gravity
darkening can be neglected. Utilizing the Fourier transform technique
\cite{Gray73, Gray76} also showed that turbulent velocities to some
extent can be distinguished from rotation.

In fast rotators the situation becomes more complex, since centrifugal
forces distort the spherical shape of the stellar surface and the line
profiles depend on the inclination the star is observed under. Beyond
the additional degree of freedom this would not be a major problem if
temperature (and pressure) were not connected to gravity. With
different regions on the stellar surface having significantly
different temperatures and gas pressures, line profiles no longer can
be approximated by convolution between a single intrinsic line profile
and the rotational broadening function.  For this reason, the effects
of fast rotation on line profiles often were studied for specific
absorption lines \citep[e.g.,][]{Stoeckley68, Hardorp68};
\cite{Collins95} come to the conclusion that the detection of
secondary effects like limb darkening and differential rotation is
improbable in line profiles of fast rotators.

Under what circumstances the approximation of various broadening
mechanisms by convolutions breaks down, depends only on the
sensitivity of the spectral line on temperature and gas pressure.
Spectral quality in terms of signal-to-noise as well as in wavelength
coverage has improved and makes possible analyses of weak absorption
features. Observers are not restricted to the extremely temperature
and pressure sensitive lines of the lightest elements H and He. Fast
rotation is the rule in stars of spectral types as late as F5, and in
the spectra of fast rotating A- and F-stars a large number of heavy
ion lines exist where the approximation of rotational broadening by a
convolution becomes valid again. Especially the deconvolution of an
``overall'' broadening function inherent in all lines by using the
methods of Least Squares Deconvolution \citep[LSD, e.g.,][]{Cameron00}
can provide reliable broadening profiles independent of line specific
intrinsic mechanisms.

In this paper the effects of fast rotation are examined in detail
focusing on geometric distortions and gravity darkening on stars
observed under different inclination angles. These effects underly all
spectral lines independent of their intrinsic shape and temperature or
pressure dependencies. The modelled spectra assume no intrinsic
temperature or pressure dependence; whether this can be applied to
specific spectral lines or groups of lines has to be checked
individually.

In \cite{Reiners02a} the possibility of detecting differential
rotation in line profiles was demonstrated. This paper also answers
the question of whether the technique used there is applicable to fast
rotators. It shows what can be learned about gravity darkening and
inclination angle from the shape of stellar absorption line profiles.

\section{Absorption profiles of fast rotators}

\subsection{Line-independent broadening}
\label{GravDarkDebate}

Absorption line profiles observed in fast rotating stars are affected
by three mechanisms inherent in all lines: (a) the projected geometry
of the star, (b) the flux distribution due to temperature variations
(known as gravity darkening) on the rotating stellar surface, and (c)
the rotation law. These three line-independent mechanisms will be
discussed in the following.

Case (a) is well understood, fast rotation diminishes the
gravitational potential at the equator and alters the star's
sphericity. As a consequence, lines of constant projected rotational
velocities no longer lie on chords but are bent \citep[cp.,
e.g.,][]{Collins95}.  Although an analytical description of the
surface velocity distribution is not then available, calculation by
integrating over the stellar surface is unproblematic.

(b) Since the presence of gravity darkening was initially described by
\cite{Zeipel24}, many publications were written on the correct gravity
dependence of surface temperature. In general, gravity darkening is
described in terms of a parameter $\beta$ used in the equation
\begin{equation}
\label{Eq:GravDarkLaw}
  T_{\rm eff} \propto g^{\beta}.
\end{equation}
The ``classical'' value of $\beta = 0.25$ found by \cite{Zeipel24} is
valid only for conservative rotation laws in stars without convective
envelopes. A value of $\beta = 0.08$ is derived for stars with
convective envelopes \citep{Lucy67}. \cite{Tuominen72} approximates a
value of $\beta \approx 0.15$ for differentially rotating stars, and
for magnetic stars without convection zones, \cite{Smith75} reports a
slightly different value of $\beta = 0.275$.  The most comprehensive
calculations were carried out by \cite{Claret98}, who provides tables
of stellar parameters including $\beta$ for a series of stars at
various evolutionary stages. The calculations are in general agreement
with the cited works; stars of spectral types later than $\sim$A2
harboring convective envelopes have a value of $\beta \approx 0.08$,
for earlier-type stars, $\beta = 0.25$ seems a good choice.  These
calculations are also consistent with measurements of $\beta$ in
eclipsing binaries carried out by \cite{Rafert80}, although a large
scatter appears in the data \citep[cp. Fig.\,7 in][]{Claret98}.

(c) Stellar rotation laws are poorly known. Our picture comes from
observations of the Sun, where the angular velocity can be
approximated as
\begin{equation}
  \label{eq:DiffRot}
  \Omega(l) = \Omega_{\rm Equator} (1-\alpha \sin^2{l}),
\end{equation}
with $l$ being the latitude and $\alpha_{\odot} \approx 0.2$, i.e.,
the Equator rotating 20\% faster than the Pole as derived from Sun
spots. Differential rotation in stars other than the Sun was claimed
for ten relatively slowly rotating F-type stars with values of
$v\,\sin{i} < 50$\,km\,s$^{-1}$ in \cite{Reiners03} adopting
Eq.\,\ref{eq:DiffRot} as the rotation law in analogy to the solar
case.  Assuming reasonable values of inclination angles $i$, the small
projected rotational velocities of the stars in that sample give no
reason to expect effects due to distortions of the gravitational
potential, and differential rotation is thought to be the cause of the
determined profile signatures. It is an open question, to what extent
differential rotation can be detected in faster rotators, where
deformation and gravity darkening become important.

\subsection{Profile shape characteristics}

Stellar line profiles are subject to a number of different mechanisms
affecting their shape. Besides the mentioned effects of gravity
darkening and rotation law, limb darkening, additional (turbulent)
velocity fields and stellar surface structure like spots can
significantly contribute to the line profiles' shape. Utilizing the
Fourier transform turns out to be of great advantage in disentangling
the different effects. When approximating line broadening by
convolutions, their interaction can be studied most easily using
Fourier transforms, since the computationally complicated convolutions
become multiplications in Fourier domain \citep[cf, e.g.,][]{Gray76}.

\cite{Reiners02a} utilized the zeros of the Fourier transformed line
profiles to search for the subtle effects of differential rotation.
They showed that the ratio of the first two zeros $q_1$ and $q_2$ is a
direct indicator for solar-like -- i.e., Equator faster that Pole --
differential rotation ($\alpha > 0$ in Eq.\,\ref{eq:DiffRot}). The
value of the ratio $q_{2}/q_{1}$ is extremely sensitive to the
profile's shape, and is unaffected by the ``standard'' turbulent
velocity fields like micro- and macroturbulence even for rotational
velocities as low as 10\,km\,s$^{-1}$. Using a linear limb darkening
law with a parameter $\epsilon$, in a rigidly rotating star the value
of $q_{2}/q_{1}$ can be approximated as
\begin{equation}
  \label{eq:q2q1_epsilon}
  \frac{q_2}{q_1} = 1.831 - 0.108\epsilon - 0.022\epsilon^2 + 0.009\epsilon^3 + 0.009\epsilon^4
\end{equation}
\citep{Dravins90}. Thus for a rigid rotator the value of $q_{2}/q_{1}$
is always between 1.72 and 1.83 regardless of the limb darkening.
Differential rotation significantly diminishes this ratio
\citep[cf][]{Reiners02a}, while polar spots lead to an increase
\citep{Reiners02b}.

Since the ratio $q_{2}/q_{1}$ turns out to be an easily accessible
parameter characteristic for the line's shape, I continue to use it
for studying the influence of gravity darkening in rapidly rotating
stars.  For a more detailed discussion of the correlations between
differential rotation $\alpha$, the inclination angle $i$ and
$q_{2}/q_{1}$ see \cite{Reiners02a, Reiners03}.

\section{Model}
The approach of this paper is to study the effects of fast rotation
universal for all absorption lines. Line specific dependencies on gas
pressure, ionization stages, etc. are not taken into account. This
approach is especially suited for studies incorporating different
lines of similar (preferably heavy) ions, e.g., with LSD methods.  For
heavy ions pressure broadening is not expected to play a major role
compared to the dominant rotational broadening. For light elements
like H and He, this is not true and great care has to be take when
applying the methods to them. Although the Lorentzian profiles induced
by pressure broadening should not add additional zeros to the line
profile Fourier transforms \citep[cp.][]{Heinzel78}, errors from
approximating line broadening effects by convolutions may become
significant in these cases.

The calculations were done using a modified version of the package
developed and described by \cite{Townsend97}. Surface integration is
carried out over 25\,500 visible surface elements scaling the flux
with temperature according to a Planck law. A Gaussian profile was
used as the input function, since the large rotational velocities make
the shape of the input function unimportant. For all cases the same
input function was used regardless of gravity and temperature. To
measure the zeros in Fourier transformed line profiles the accuracy
must be higher for the faster rotators; 32\,768 points were used in
the transform algorithm for values of $v\,\sin{i} \le
100$\,km\,s$^{-1}$, 65\,536 points for values of $v\,\sin{i}$ between
100 and 200\,km\,s$^{-1}$, and 131\,072 points for $v\,\sin{i} >
200$\,km\,s$^{-1}$. Thus the sampling error in $q_{2}/q_{1}$ does not
exceed a value of 0.01 for all calculations.

Broadening profiles are calculated and the values of $q_{2}/q_{1}$
determined for four stellar models with polar effective temperature
$T_{\rm eff}^{\rm Pole}$ and mass given in columns two and three in
Tab.\,\ref{tab:Coefficients}. A grid is computed in three values of
$\beta$ (0.0, 0.08 and 0.25), five values of inclination angle $i$
(10$^{\circ}$, 30$^{\circ}$, 50$^{\circ}$, 70$^{\circ}$ and
90$^{\circ}$), and values of $v\,\sin{i}$ between 50 and
300\,km\,s$^{-1}$ in steps of 25\,km\,s$^{-1}$ with the equatorial
velocity $v_{\rm e} < 350$\,km\,s$^{-1}$. Altogether more than 450
models are computed. The dependence of the ratio $q_{2}/q_{1}$ on
different parameter choices is analyzed and studied in the following.

\section{Line shape variations}

\begin{table}
  \caption{\label{tab:Coefficients} Parameters of the calculated models,
    $a$ and $b$ are parameters used in Eq.\,\ref{eq:Polynomial} with values of 
    gravity darkening $\beta$ according to \cite{Claret98}. Note that
    for all models profiles with $\beta = 0.0, 0.08$ and $0.25$ were analyzed}
  \begin{tabular}{crrccc}
    \hline
    \hline
    \noalign{\smallskip}
    Model & $T_{\rm eff}^{\rm Pole}$ & $M/M_\odot$ & $\beta$ & $a$ & $b$\\
    \noalign{\smallskip}
    \hline
    \noalign{\smallskip}
    B0 & 30\,000 & 17.5 & 0.25 & .740E-4 & -.345E-6\\
    A0 & 10\,000 & 2.4 & 0.25 & .190E-3 & -.136E-5\\
    F0 &  7\,000 & 1.6 & 0.08 & .172E-3 & -.993E-6\\
    G0 &  5\,700 & 1.0 & 0.08 & .184E-3 & -.116E-5\\
    \noalign{\smallskip}
    \hline
  \end{tabular}
\end{table}

\subsection{Monotonic dependence on rotational velocity $v_{\rm e}$}

\begin{figure}
  \centering
  \resizebox{\hsize}{!}{\includegraphics[angle=-90]{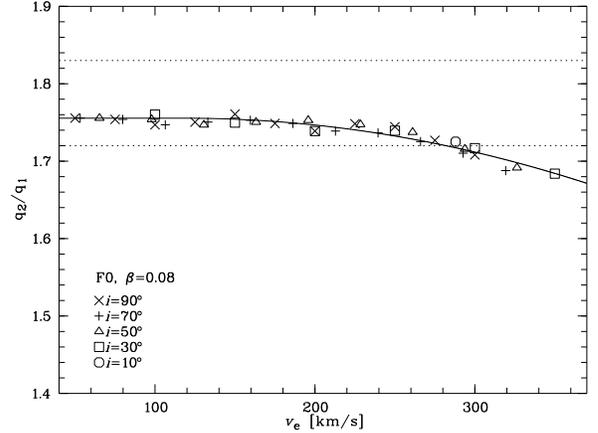}}
  \caption{\label{plot:F0_vsini} Derived values of $q_{2}/q_{1}$ for a
    rigidly rotating F0-type star with $\beta = 0.08$. Different
    symbols indicate inclination angles as indicated in the figure.
    The value of $q_{2}/q_{1}$ is determined by $v_{\rm e}$
    independently of the values of $i$ and $v\,\sin{i}$. The slope is
    well approximated by the second order polynomial shown as a solid
    line.}
\end{figure}

To show the dependence of $q_{2}/q_{1}$ on $v_{\rm e}$, in
Fig.\,\ref{plot:F0_vsini} $q_{2}/q_{1}$ is plotted vs.  $v_{\rm e}$
for the F0 model with a gravity darkening parameter of $\beta = 0.08$.
The region in $q_{2}/q_{1}$ typical for a spherical surface and a
linear limb darkening law with $0.0 < \epsilon < 1.0$ is indicated
with dotted lines (cp. Eq.\,\ref{eq:q2q1_epsilon}). Values for stars
seen under various inclination angles are distinguished by different
symbols, all values are in good agreement with the marked polynomial
indicating a smooth decline of $q_{2}/q_{1}$ for stars rotating faster
than $v_{\rm e} \approx 150$\,km\,s$^{-1}$.

The parameter determining the value of $q_{2}/q_{1}$ in a given
stellar model obviously is $v_{\rm e}$, independent of the inclination
angle (and thus $v\,\sin{i}$). This means, that under the assumption
of rigid rotation, and with assumed values of $\beta$, in fast
rotators one can in principle obtain the value of $v_{\rm e}$ from the
value of $q_{2}/q_{1}$, and thus, with the known value of
$v\,\sin{i}$, the inclination angle $i$.  I will return to this point
in Sect.\,\ref{Sect:Situation}.

\subsection{Gravity darkening and spectral class}

\begin{figure*}
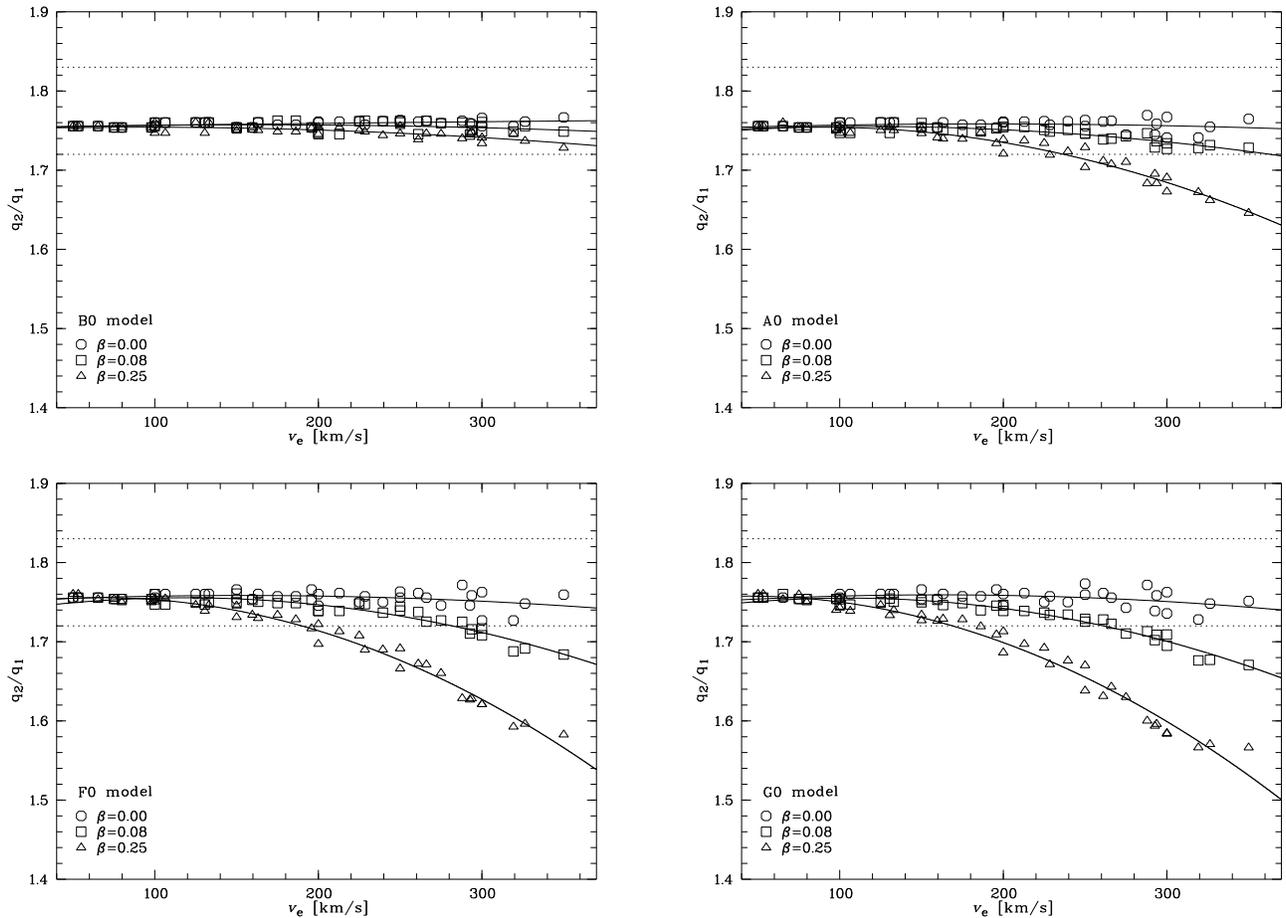

  \centering \mbox{
    \resizebox{.5\hsize}{!}{\includegraphics[angle=-90]{3947.f2a}}
    \resizebox{.5\hsize}{!}{\includegraphics[angle=-90]{3947.f2b}}}
  \mbox{ \resizebox{.5\hsize}{!}{\includegraphics[angle=-90]{3947.f2c}}
    \resizebox{.5\hsize}{!}{\includegraphics[angle=-90]{3947.f2d}}}
  \caption{\label{plot:stars}Derived values of $q_{2}/q_{1}$ vs. equatorial
    velocity $v_{\rm e}$ for four different stellar models (B0, A0, F0
    and G0). For each model different values of the gravity
    darkening parameter $\beta$ (0.0, 0.08 and 0.25) are indicated
    using different symbols. The dependence of $q_{2}/q_{1}$ on
    $v_{\rm e}$ differs for the different cases but is monotonic in
    $\beta$ and $v_{\rm e}$. }
\end{figure*}

In the four panels of Fig.\,\ref{plot:stars} the derived values of
$q_{2}/q_{1}$ are plotted vs. $v_{\rm e}$ for the models of spectral
types B0, A0, F0 and G0 as specified in Tab.\,\ref{tab:Coefficients}.
In each panel calculations for three different values of $\beta$ are
shown, the different choices of $\beta$ (0.0, 0.08 and 0.25) are
plotted using different symbols. The slope of $q_{2}/q_{1}$ is well
described by a second order polynomial for all cases. The ratio
$q_{2}/q_{1}$ is diminished in rapid rotators with larger deviations
from the standard value of $q_{2}/q_{1} = 1.75$ the larger the value
of $\beta$.  In the A0, F0 and G0 models, $q_{2}/q_{1}$ becomes less
than 1.72 for very fast rotators depending on the value of $\beta$,
i.e., deviations from standard line profiles are significant and in
principle observable in these stars.

The behavior of $q_{2}/q_{1}$ with different values of the gravity
darkening parameter $\beta$ can be studied in the four panels of
Fig.\,\ref{plot:stars}. In the B0 model no value of $q_{2}/q_{1} <
1.72$ appears, but the influence of $\beta$ on the ratio $q_{2}/q_{1}$
grows with later spectral type. In all cases $q_{2}/q_{1}$ depends
monotonically on the rotational velocity $v_{\rm e}$ with diminished
$q_{2}/q_{1}$ for larger rotational velocities. The result that the
line profiles depend more strongly on the value of $v_{\rm e}$ for
stars of later spectral types, reflects the fact that the relative
temperature contrast $\Delta T/T$ on the stellar surface is larger for
cooler stars, since $\Delta T$ is comparable for similar values of
$v_{\rm e}$. Furthermore, for all models no significant variation of
$q_{2}/q_{1}$ appears for $\beta = 0$, i.e., without gravity darkening
the geometrical deformation does not suffice to significantly change
the shape of the line profiles.

In a larger sample of stars of similar spectral types these results
could make it feasible to determine the value of gravity darkening in
terms of the parameter $\beta$ when plotting $q_{2}/q_{1}$ vs.
$v\,\sin{i}$. While for many stars in such a plot the values should
fall below an upper envelope expected from the $q_{2}/q_{1}$--$v_{\rm
  e}$ relation (the real values of $v_{\rm e}$ are larger than the
measured values of $v\,\sin{i}$), a clear threshold should mark the
upper envelope indicating the correct value of $\beta$. Using a larger
sample of fast rotators of similar spectral types the ratio
$q_{2}/q_{1}$ could thus be used to determine the values of $\beta$
from measurements of single stars.

\subsection{Determination of $i$}
\label{Sect:Situation}

\begin{figure*}
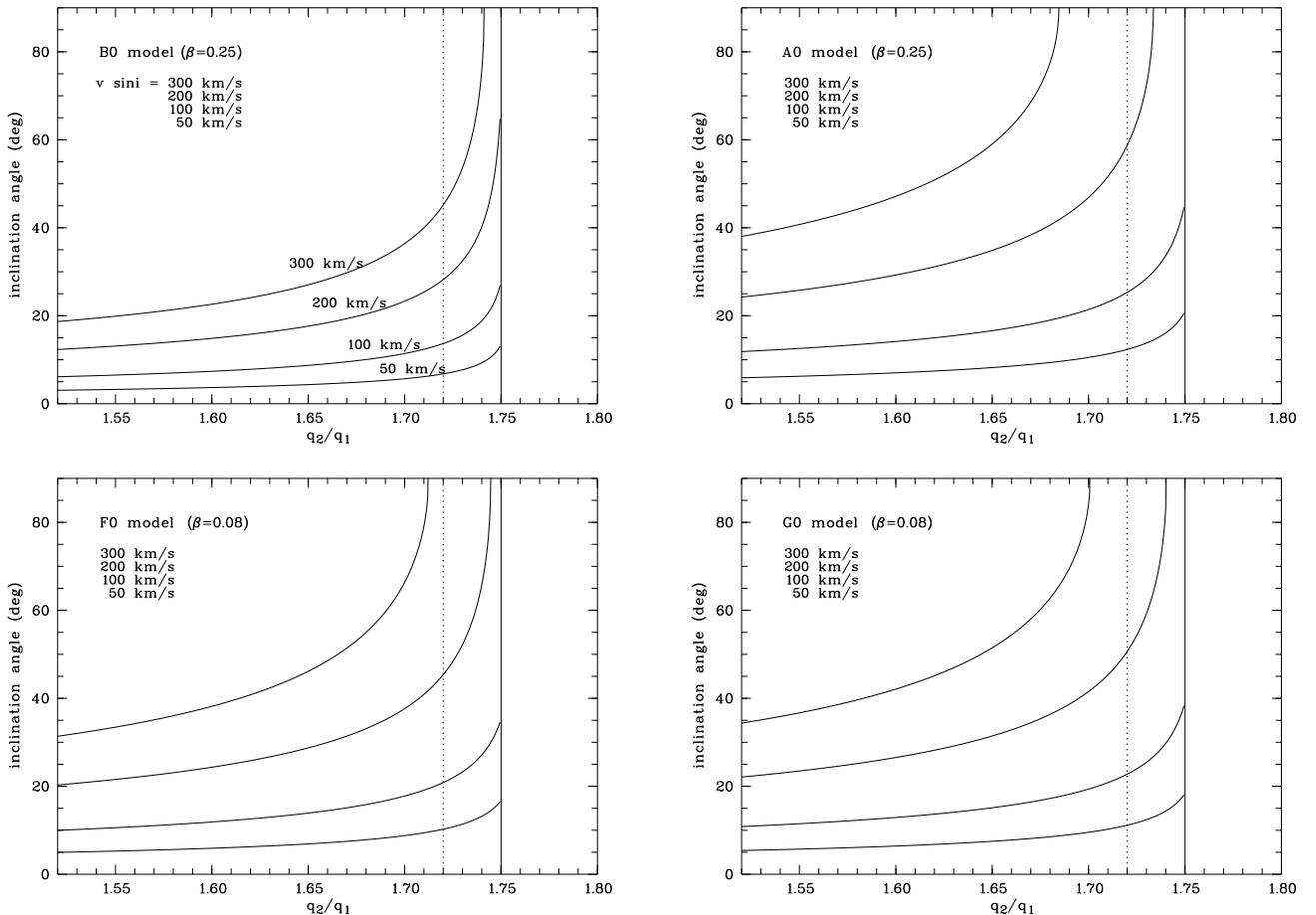

  \centering \mbox{
    \resizebox{.5\hsize}{!}{\includegraphics[angle=-90]{3947.f3a}}
    \resizebox{.5\hsize}{!}{\includegraphics[angle=-90]{3947.f3b}}}
  \mbox{
    \resizebox{.5\hsize}{!}{\includegraphics[angle=-90]{3947.f3c}}
    \resizebox{.5\hsize}{!}{\includegraphics[angle=-90]{3947.f3d}}}
  \caption{\label{plot:i_q2q1}Stars with measured values of $v\,\sin{i}$ and
    $q_{2}/q_{1}$ are seen under an inclination angle $i$. In the four
    panels (B0, A0, F0 and G0-type stars) the approximated dependence
    of $i$ on the ratio $q_{2}/q_{1}$ is shown for values of
    $v\,\sin{i} = 50, 100, 200$ and $300$\,km\,s$^{-1}$ (solid lines,
    top to bottom as indicated in the upper left panel). The dashed
    line marks the lowest ratio of $q_{2}/q_{1}$ consistent with slow
    rotation and extreme limb darkening, no values of $q_{2}/q_{1} >
    1.75$ are expected.}
\end{figure*}

The ratio $q_{2}/q_{1}$ depends monotonically on equatorial velocity
$v_{\rm e}$ and is independent of the inclination angle $i$. Assuming
a value of $\beta$ for a fast rotator of known spectral class, the
equatorial velocity $v_{\rm e}$ can thus be determined from a
measurement of $q_{2}/q_{1}$. The functions $q_{2}/q_{1}(v_{\rm e})$
shown in Fig.\,\ref{plot:stars} are approximated by the second order
polynomial
\begin{equation}
  \label{eq:Polynomial}
  q_{2}/q_{1} = 1.75 + av_{\rm e} + bv_{\rm e}^2.
\end{equation}
The parameters $a$ and $b$ for the four models using the most probable
values of the parameter $\beta$ according to \cite{Claret98} are given
in Tab.\,\ref{tab:Coefficients}. With Eq.\,\ref{eq:Polynomial} it is
possible to derive the inclination angle $i$ from a standard
measurement of $v\,\sin{i}$ and the value of $q_{2}/q_{1}$.  

In each panel of Fig.\,\ref{plot:i_q2q1} the values of $i$ are plotted
vs. the ratio $q_{2}/q_{1}$ for four different values of $v\,\sin{i}$.
As in Fig.\,\ref{plot:stars} each panel represents a stellar model
from Tab.\,\ref{tab:Coefficients}. Functions for values of $v\,\sin{i}
= 50, 100, 200$ and $300$\,km\,s$^{-1}$ are plotted as examples; for
calculating intermediate values of $v\,\sin{i}$,
Eq.\,\ref{eq:Polynomial} can be used with values of $a$ and $b$ as
given in columns five and six of Tab.\,\ref{tab:Coefficients}. For
example, a value of $q_{2}/q_{1} = 1.70$ measured in a profile of an
A0-type star with $v\,\sin{i} = 100$\,km\,s$^{-1}$ can indicate that a
rigidly rotating star is observed under an inclination of $i =
20^{\circ}$ and thus $v_{\rm e} \approx 290$\,km\,s$^{-1}$.

\subsection{Differential rotation in fast rotators}

Solar-like differential rotation with the Equator faster than the Pole
diminishes the value of $q_{2}/q_{1}$ as shown in \cite{Reiners02a}; a
measurement yielding $q_{2}/q_{1} < 1.72$ can be due to a solar-like
rotation law with the Equator faster than the Pole. Whether very fast
rotators are subject to strong differential rotation will not be
discussed here. In Fig.\,\ref{plot:stars} $q_{2}/q_{1}$ appears to be
larger than 1.72 for all stars with $v_{\rm e} < 200$\,km\,s$^{-1}$
and values of gravity darkening according to \cite{Claret98} (cp.
Tab.\,\ref{tab:Coefficients}). This means that a measurement of
differential rotation by determining the ratio $q_{2}/q_{1}$ is not
affected by gravity darkening if $v_{\rm e} \la 200$\,km\,s$^{-1}$.
Interpreting gravity darkening as the reason for a measured value of
$q_{2}/q_{1} < 1.72$, the rotational velocity ($v_{\rm e} >
200$km\,s$^{-1}$) and inclination angle of the star can be determined.
Consistency of $v_{\rm e}$ with breakup velocity can be checked. For
$v_{\rm e} < v_{\rm crit}$, differential rotation and very fast
rotation remain indistinguishable without further information. An
example is given in Sect.\,\ref{sect:SampleExample}.

\section{Real and measured values of $\mathbf{v}$\,sin$\mathbf{i}$}

\begin{figure*}
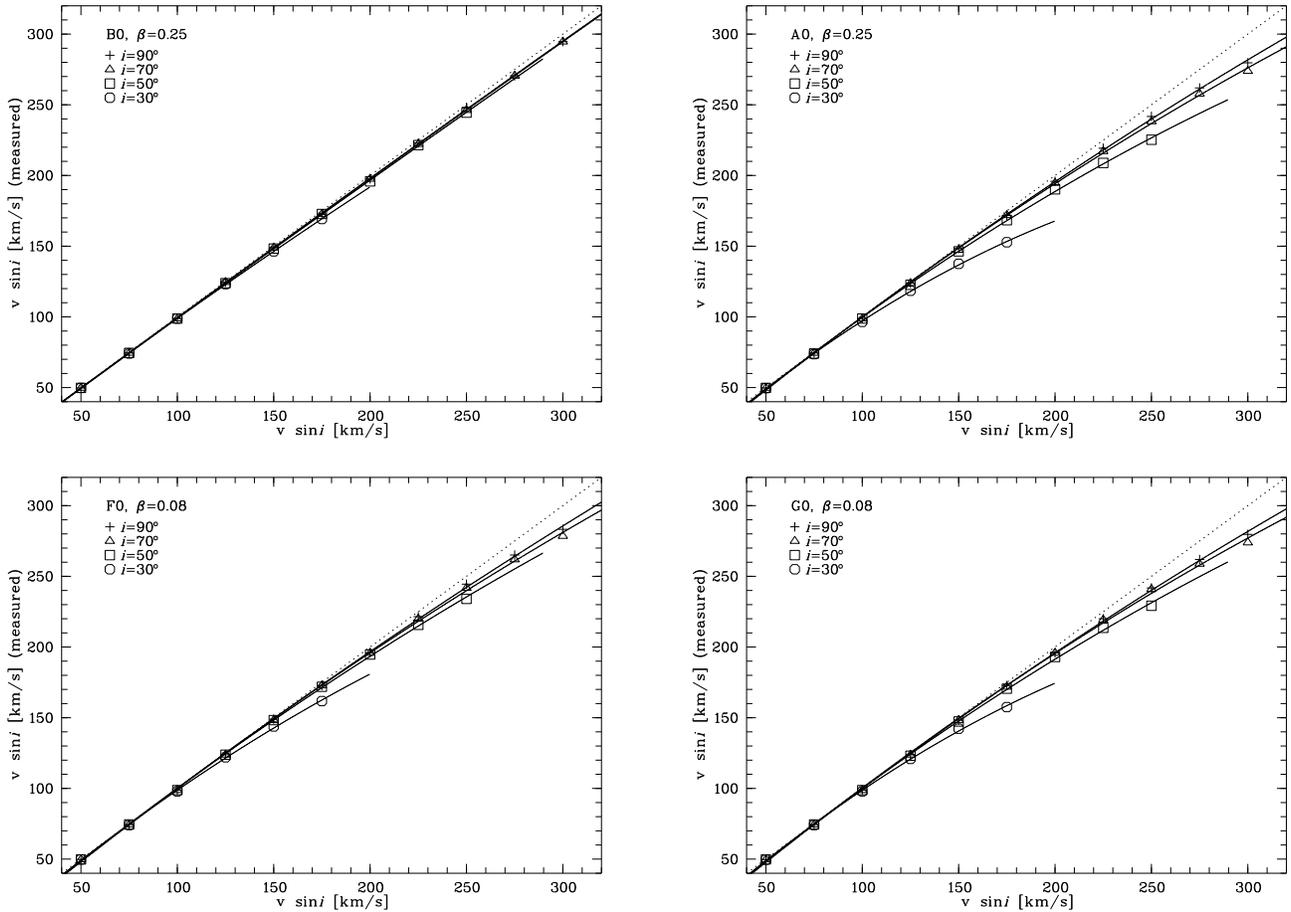

  \centering
  \mbox{
  \resizebox{.5\hsize}{!}{\includegraphics[angle=-90]{3947.f4a}}
  \resizebox{.5\hsize}{!}{\includegraphics[angle=-90]{3947.f4b}}}
\mbox{
  \resizebox{.5\hsize}{!}{\includegraphics[angle=-90]{3947.f4c}}
  \resizebox{.5\hsize}{!}{\includegraphics[angle=-90]{3947.f4d}}}
  \caption{  \label{plot:vsini}Values of $v\,\sin{i}$ as measured from
    the first zero of the Fourier transformed broadening profiles
    plotted vs. the real values given as input to the models. The four
    panels show cases of B0, A0, F0 and G0-type stars, the value of
    the gravity darkening parameters $\beta$ are chosen according to
    \cite{Claret98}. In each panel series for four different
    inclination angles ($i=90^{\circ}, 70^{\circ}, 50^{\circ}$ and
    $30^{\circ}$) are differentiated, the respective values are
    connected by solid lines. Dotted lines mark the identity between
    real and measured values.}
\end{figure*}

To what accuracy a measurement of $v\,\sin{i}$ is possible was
discussed in a number of publications \cite[e.g.,][]{Stoeckley68,
  Collins95}. It is known that measured values of $v\,\sin{i}$
underestimate the real values especially for very fast rotators
regardless of the data quality. To what extent a measured value is too
low, depends on the gravity darkening law and on the stellar
inclination. In the four panels of Fig.\,\ref{plot:vsini} the
differences between real and measured values for the B0, A0, F0 and
G0-type stars are shown using the most probable values of gravity
darkening as done above. The measured values of $v\,\sin{i}$ have been
derived using the first zeros of the Fourier transform. Values for
different inclination angles are differentiated and shown with
different symbols, those derived for identical values of $i$ are
connected by solid lines. Dashed lines mark the identity of real and
measured values.  As expected, the measured values of $v\,\sin{i}$ may
underestimate the real ones but will never indicate too high a
projected rotational velocity. This reflects the fact that the
(equatorial) surface regions with the largest projected rotational
velocities become cooler due to gravity darkening, and thus less flux
is observed from them.

The extent, to which a measured value of $v\,\sin{i}$ is
systematically underestimated, depends strongly and in a well-defined
manner on the inclination angle. For constant inclination angles, the
measured value of $v\,\sin{i}$ is a smooth function of the real one
\citep[cp.  to Fig.\,6 in][]{Collins95}.  Underestimation of
$v\,\sin{i}$ is stronger for smaller inclination angles in all models.
The deviation naturally is higher for large values of $v\,\sin{i}$
(and thus also of $v_{\rm e}$). Thus, measured values of $v\,\sin{i}$
in very rapid rotators can be corrected by determining the ratio
$q_{\rm 2}/q_{\rm 1}$ and calculating the inclination angle $i$ as
shown in Sect.\,\ref{Sect:Situation}.

It should be noted that -- in contrast to Fig.\,12 in \cite{Collins95}
-- inclination has a strong effect on the zeros of Fourier transformed
line profiles of rapid rotators; the case shown there only holds for
zero projected rotational velocity.

\section{Example: Differential or fast rotation?}
\label{sect:SampleExample}

\begin{table}
  \caption{\label{tab:DiffStars}Stars with measured values $q_{2}/q_{1} < 1.72$ from the sample of
    \cite{Reiners03}; differential rotation is claimed for these objects.
    Interpreting the values of $q_{2}/q_{1}$ as due to very fast rotation
    seen under small inclination angles, the required values of $v_{\rm e,
      rigid}$ and $i$ are given. Nine of the ten are larger than break-up
    velocity.}
  \begin{tabular}{rlcccc}
    \hline
    \noalign{\smallskip}
    HD & Type & $v\,\sin{i}$& $q_{2}/q_{1}$ & $v_{\rm e, rigid}$ & $i$\\
    &&[km\,s$^{-1}$]&&[km\,s$^{-1}$]\\  
    \noalign{\smallskip}
    \hline
    \noalign{\smallskip}
    89449 & F6IV & $17.3 \pm 1.7$ & $1.44$ & 630 & 1.7\degr\\
    89569 & F6V & $12.2 \pm 0.7$ & $1.57$ & 500 & 1.4\degr\\
    100563 & F5V & $13.5 \pm 0.4$ & $1.67$ & 370 & 2.1\degr\\
    105452 & F0IV/V & $23.5 \pm 1.2$ & $1.59$ & 500 & 2.7\degr\\
    120136 & F7V & $15.6 \pm 1.0$ & $1.57$ & 500 & 1.8\degr\\
    121370 & G0IV & $13.5 \pm 1.3$ & $1.46$ & 590 & 1.3\degr\\
    160915 & F6V & $12.4 \pm 0.5$ & $1.60$ & 470 & 1.5\degr\\
    173667 & F6V & $18.0 \pm 2.0$ & $1.40$ & 660 & 1.6\degr\\
    175317 & F5IV/V & $17.1 \pm 0.7$ & $1.58$ & 490 & 2.0\degr\\
    197692 & F5V & $41.7 \pm 1.7$ & $1.62$ & 440 & 5.4\degr\\
    \hline
  \end{tabular}
\end{table}

As an observational example the ten stars for which differential
rotation was claimed in \cite{Reiners03} are given in
Tab.\,\ref{tab:DiffStars} with their determined values of $v\,\sin{i}$
and $q_{2}/q_{1}$.  According to the present calculations, these
values of $q_{2}/q_{1} < 1.72$ could also be interpreted as fast
rotation seen under small inclination angles $i$. The requested values
of $v_{\rm e}$ and $i$ for this interpretation are calculated and
given in columns five and six in Tab.\,\ref{tab:DiffStars}. With one
exception all values of $v_{\rm e}$ are larger than break-up velocity
and thus can be excluded (it is furthermore very unlikely to find
inclination angles below 6\degr\ in ten out of the 32 objects used in
that sample). Thus differential rotation remains the most probable
explanation for the peculiar shapes of these profiles.

\section{Summary}

Line independent effects of fast stellar rotation on line profiles
were presented. Neglecting line specific variations in shape and line
depth, the effects of geometrical deformation and the role of the
gravity darkening parameter $\beta$ have been outlined. Grid
calculations in gravity darkening $\beta$, inclination angle $i$,
rotational velocity $v_{\rm e}$ and spectral type with the parameter
$q_{2}/q_{1}$ classifying the profiles' shapes were carried out. The
results summarized below can be applied especially to high quality
data covering large wavelength regions where line-independent
broadening profiles may be obtained using the methods of LSD.
Applicability must be proved for each case individually, but
especially for lines of heavier ions in stars of later spectral types,
intrinsic variations of line profiles are sufficiently small and the
results of this study can directly be applied. The results can be
summarized as follows:

\begin{enumerate}
\item In very fast rotators ($v\,\sin{i} > 150$\,km\,s$^{-1}$) and for
  constant values of $\beta$ and similar spectral types, the value of
  $q_{2}/q_{1}$ shows monotonic dependence on equatorial velocity
  $v_{\rm e}$ with $q_{2}/q_{1}$ diminishing significantly with larger
  values of $v_{\rm e}$. Variations with inclination $i$ are marginal.
\item The coupling of $q_{2}/q_{1}$ on $v_{\rm e}$ depends on the
  value of the gravity darkening parameter $\beta$ and of spectral
  type. Larger values of $\beta$ lead to lower values of $q_{2}/q_{1}$
  with stronger slopes for later spectral types.  Estimating the
  physical values of $\beta$ according to \cite{Claret98}, the
  parameter $q_{2}/q_{1}$ becomes significantly diminished in fast
  rotating ($v_{\rm e} \ga 200$\,km\,s$^{-1}$) A-, F- and G-type
  stars.  Independent of the inclination angle $i$, the equatorial
  velocity $v_{\rm e}$ can be estimated by measuring $q_{2}/q_{1}$,
  and thus measuring $v\,\sin{i}$ and $v_{\rm e}(q_{2}/q_{1})$
  determines the inclination angle $i$.
\item For stars with rotational velocities of $v_{\rm e} \la
  200$\,km\,s$^{-1}$ the value of the ratio $q_{2}/q_{1}$ does not
  significantly change and is larger than 1.72 for all choices of
  $\beta$ and for all spectral types.
\item For fixed values of $v\,\sin{i}$ the zeros of Fourier
  transformed rotational broadening profiles do depend on inclination
  angle $i$. A smaller value of $i$ results in a larger difference
  between the value of $v\,\sin{i}$ as measured from the first zero of
  the Fourier transform and its real value. Measured projected
  rotational velocities are never overestimated. Deviations are as
  large as 20\% depending on the value of $i$.  Using the parameter
  $q_{2}/q_{1}$ this effect can be corrected for.
\end{enumerate}

As for differential rotation, it turned out that its measurement is
not affected by gravity darkening in stars rotating slower than
$v_{\rm e} \approx 200$\,km\,s$^{-1}$. For faster rotators, small
values of $q_{2}/q_{1}$ can principally be due to three reasons; (i)
strong differential rotation, (ii) very fast rigid rotation (perhaps
seen under a small inclination angle), and (iii) both. The constraints
on a value of $v_{\rm e}(q_{2}/q_{1})$ can help to clarify the
situation, e.g., if a value of $v_{\rm e}$ larger than breakup
velocity is needed to rule out differential rotation.

\begin{acknowledgements}
  A.R. acknowledges financial support from Deutsche
  Forschungsgemeinschaft DFG-SCHM 1032/10-1.
\end{acknowledgements}

\end{document}